\documentclass[]{rsos}
\usepackage{graphicx}
\usepackage{xcolor}

\titlehead{Research}
\begin{document}
\title{Generic catastrophic poverty when selfish
investors exploit a degradable common resource}
\author{Claudius Gros}
\address{Institute for Theoretical Physics, Goethe
University Frankfurt, Germany} 
\subject{systems theory, mathematical modeling}

\corres{Claudius Gros\\
\email{gros07[@]itp.uni-frankfurt.de}}

\begin{abstract} 
The productivity of a common pool of resources may 
degrade when overly exploited by a number of selfish 
investors, a situation known as the tragedy of the 
commons (TOC). Without regulations, agents optimize
the size of their individual investments into the commons
by balancing incurring costs with the returns received. 
The resulting Nash equilibrium involves a self-consistency 
loop between individual investment decisions and 
the state of the commons. As a consequence, several 
non-trivial properties emerge. For $N$ investing actors 
we prove rigorously that typical payoffs do not scale 
as $1/N$, the expected result for cooperating agents, 
but as $(1/N)^2$. Payoffs are hence reduced with
regard to the functional dependence on $N$,
a situation denoted catastrophic poverty.
We show that catastrophic poverty results from a
fine-tuned balance between returns and costs. 
Additionally, a finite number of oligarchs may 
be present. Oligarchs are 
characterized by payoffs that are finite and 
not decreasing when $N$ increases. Our results 
hold for generic classes of models,
including convex and moderately concave cost
functions. For strongly concave cost functions
the Nash equilibrium undergoes a collective 
reorganization, being characterized instead 
by entry barriers and sudden death forced 
market exits. 
\end{abstract}

\begin{fmtext}   
\end{fmtext}   
\maketitle

\section{Introduction}

The tragedy of the commons (TOC) \cite{rankin2007tragedy}
occurs when a common resource, the commons,
is overly exploited by a number of selfish agents 
\cite{frischmann2019retrospectives}. Unmanaged access to the 
commons allows for the maximization of individual 
profits, which may lead in turn to a potentially severe
reduction of overall welfare. 
Given the essential
importance of land, water and other environmental 
resources, it may not surprise that human 
societies developed over time a large array of 
regulative options for their preservation 
\cite{ostrom1990governing,feeny1990tragedy}.
On a global level it is however not yet clear
whether humanity's present course of actions 
will, or will not lead to a world-wide tragedy of the 
commons \cite{cohen1995population,battersby2017news}.
Over-exploitation can be avoided from a game-theoretical 
perspective when appropriate rules are introduced
\cite{faysse2005coping}. In this context, the role
of evolving strategies \cite{hilbe2013evolution,stewart2014collapse}
has been discussed, as well as situations that can be 
modeled using cooperation- or coordination games
\cite{battersby2017news, carrozzo2021tragedy}.
Analogous questions arise with regard to the 
interplay between the utilization of common resources 
and Darwinian competition within natural habitats
\cite{killingback2006evolution,gore2009snowdrift,kummerli2010molecular}.
On a social level, conflicting objectives between individual 
and group interests are ubiquitous \cite{milinski2008collective},
f.i.\ when it comes to cope with climate
changes \cite{capstick2013public}, and in the context of
vaccination campaigns \cite{bauch2003group,galvani2007long}.
With regards to the time domain, it has been shown
that the dynamics of the feedback loop between 
environmental degradation and individual 
investment decisions may lead to a large variety 
of complex dynamical states
\cite{brown2007durability,weitz2016oscillating,tilman2020evolutionary}.

A wide range of situations involve the tragedy
of the commons \cite{dietz2003struggle}. In 
a  basic framework a group of agents invests
independently into a common resource \cite{ostrom1999coping}. 
The returns received are however dependent on the 
status of the commons, which degrades as a function 
of overall investment. A reference example 
is the depletion of an extended underground 
aquifer when the distributed extraction of 
water remains unchecked. Without coordination, 
pumping continues as long as it is economically 
viable. The size of the profit made is not directly 
relevant in this situation, as long as it remains 
positive. Falling levels of the water reservoir
will require however deeper and more powerful 
wells, which is equivalent to a decreasing 
productivity of the commons.

Here we will not investigate how the tragedy
of the commons may be averted. Instead, we
examine in detail the TOC equilibrium, namely
the steady state resulting from individual 
profit maximization. In this state the degree 
of exploitation remains finite for all numbers
of participating agents, becoming potentially 
large only for small investment costs. Intuitively 
one may expect that agents receive individual 
payoffs of the order of $1/N$, where $N$ is the 
number of participating agents. This is however 
not the case. We prove that payoffs scale as $(1/N)^2$ 
for the majority of agents. A scaling with $(1/N)^2$ 
entails that payoffs are dramatically reduced 
when $N$ is large, a situation denoted `catastrophic 
poverty'. Using numerical simulations, we find
that this effect becomes relevant already
for $N\sim10-20$. In addition to the agents suffering
from catastrophic poverty, there may exist
a finite number of investors with substantial 
profits, the oligarchs. In the TOC equilibrium 
the group of oligarchs is finite in the sense 
that their number cannot be proportional to $N$.

An analogy can be made to the optimization 
of production in an elastic market 
\cite{nicholson2012microeconomic}. Firms 
produce goods at a factor price,
the per-unit cost, for which they receive 
a market price that decreases when the 
total number of goods produced is increased.
Market prices are in this analogy identical 
for all producers, but not the individual 
cost functions. We prove that profits are 
quadratic and not linear functions 
of per-unit costs when firms optimize their 
outputs individually. This observation holds
in analogy for the TOC equilibrium.

In a first step, a basic reference model is 
used for the development of the framework.
Subsequently we show that catastrophic poverty 
arises generically when marginal costs are 
either constant, increasing or moderately decreasing 
when the investment into the commons is increased. 
The nature of the Nash equilibrium changes however 
qualitatively when economies of scale 
become pronounced, which corresponds to the
case of strongly concave costs functions.
In this regime catastrophic poverty is
absent. Instead, new investors with higher 
costs will face a barrier to enter 
the market. The entry barrier arises because
equilibrium investment levels are governed
by a saddle-node bifurcation, which changes also 
the timescale of forced-market exits.

\begin{figure}[!t]
\centerline{
\framebox{\color{white} \framebox{
\includegraphics[width=0.75\columnwidth]{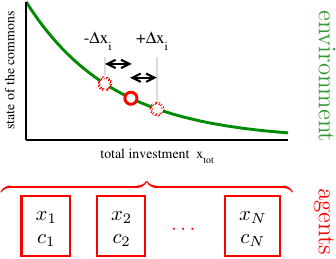}
}}
           }
\caption{{\bf Modeling framework.} Agents contribute $x_i$
to the total investment $x_{\rm tot}=\sum_i x_i$,
which determines in turn the degree of
exploitation of the commons (`the state'). 
Productivity (green line) decreases with raising 
levels of exploitation, reacting in response
to a change by $\pm\Delta x_i$ of an individual 
contribution. Payoffs per investment are
given by the productivity of the commons 
minus the individual per-unit costs $c_i$.
\label{fig_illustration}
}
\end{figure}

\section{Reference model for the tragedy of the commons}

In a first step we use a basic reference model for 
the development of the discussion. We will keep the
model as simple as possible, showing subsequently 
that the conclusions reached hold for frameworks 
generalized along several directions. 

Agents receive payoffs $E_i$ from investing
amounts $x_i\ge0$ into the commons. Here
$i=1,..,N$, with $N$ being the number of
participating agents. Investments of all
sizes are allowed as long as payoffs remain
positive. The payoffs are given by
the difference between the nominal return
and investment costs,
\begin{equation}
E_i = \left(\mathrm{e}^{-x_{\rm tot}}-c_i\right)x_i,
\qquad\quad
x_{\rm tot} = \sum_j x_j\,,
\label{TOC_E_i}
\end{equation}
where $c_i>0$ is the per-unit cost of the agent. 
Marginal costs are assumed here to be constant, 
a specification that will be relaxed further on. 
The factor $\exp(-x_{\rm tot})$ specifies how the 
productivity of the commons (the nominal return 
per unit investment) decreases as a function 
of total investments $x_{\rm tot}$. Alternative 
functional forms for the decay of the productivity
with increasing exploitation will be discussed 
further down together with a close relation of 
(\ref{TOC_E_i}) to an experimental protocol 
used by Ostrom \cite{ostrom1999coping}.
A graphical illustration of our
approach is given in Fig.~\ref{fig_illustration}.
  
\subsection{Finite productivity}

\begin{figure}[!t]
\centerline{
\includegraphics[width=0.75\columnwidth]{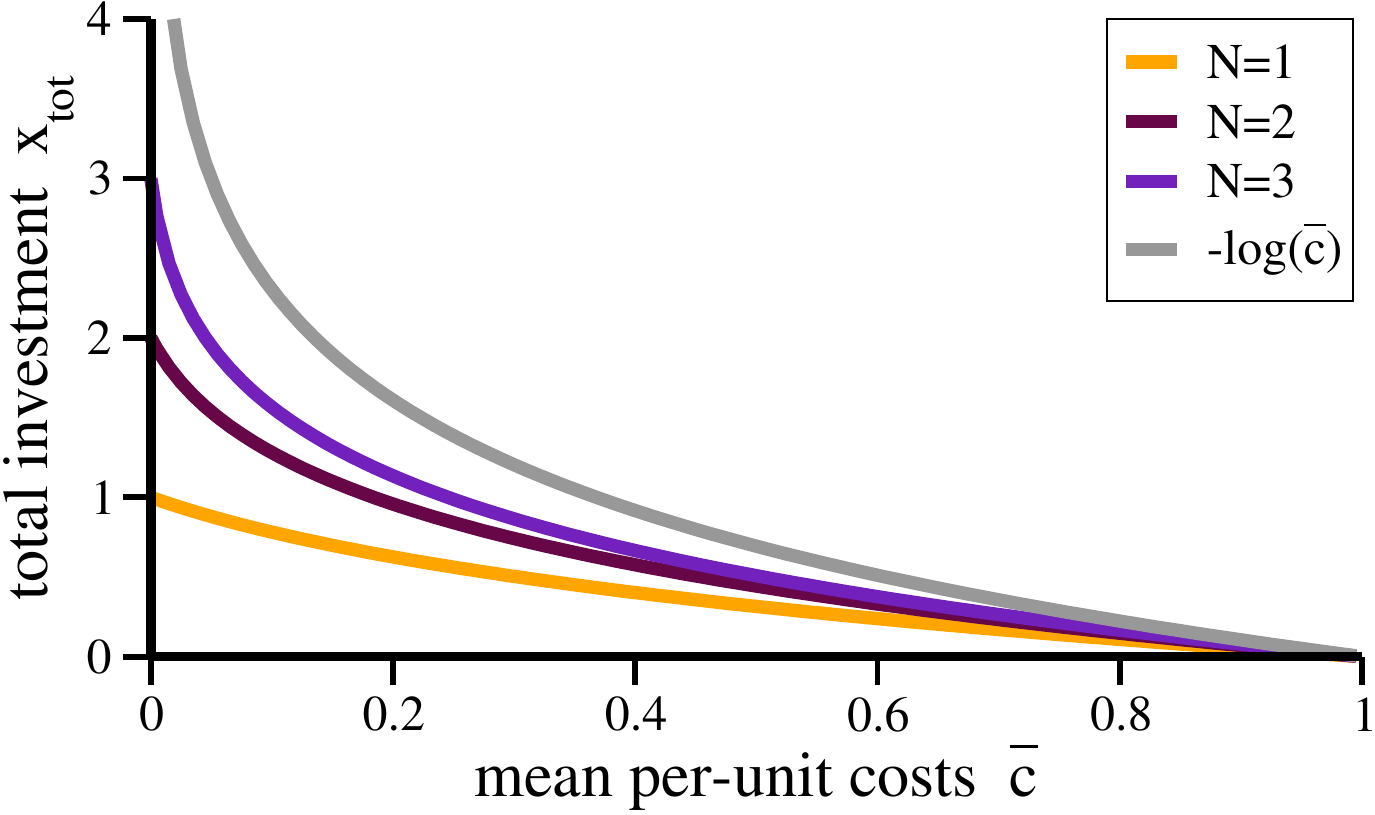}
           }
\caption{{\bf Optimal total investments.}
For various numbers of agents, $N$, the
solution of the self-consistency condition
for $x_{\rm tot}$, Eq.~(\ref{TOC_x_tot_optimal}).
The data interpolates smoothly between 
$x_{\rm tot}=N$ at $\bar{c}=0$ and
$x_{\rm tot}=0$ for $\bar{c}=1$. 
For $N\to\infty$ one has
$x_{\rm tot}=\log(1/\bar{c})$, which diverges
logarithmically for vanishing average costs, 
$\bar{c}\to0$.
\label{fig_x_tot_c_bar}
}
\end{figure}

Formally, the common pool of resources
defined with (\ref{TOC_E_i}) has an infinite
size, in the sense that $x_i$ can be 
arbitrary large. The effective productivity 
of the commons is nevertheless limited. This 
is evident when considering the case of a 
single investor, $N=1$, setting $x=x_i$ and 
$c=c_i$. The payoff is 
\begin{equation}
E = x\left(\mathrm{e}^{-x}-c\right)\,,
\label{TOC_E_i_one}
\end{equation}
which becomes negative when $\exp(-x)<c$. 
Optimality is reached when $dE/dx=0$, which
leads to
\begin{equation}
(1-x)\mathrm{e}^{-x} = c,
\qquad\quad
E\big|_{\rm opt} = x^2\mathrm{e}^{-x}\,.
\label{TOC_E_tot_one_optimal}
\end{equation}
The solution of the self-consistency condition 
for $x$, the first equation in
(\ref{TOC_E_tot_one_optimal}), leads to 
a finite optimal investment, $0\le x\le1$,
which interpolates smoothly between
$x=0$ (for $c=1$) and $x=1$ (for $c=0$), 
as shown in Fig.~\ref{fig_x_tot_c_bar}.
The decaying productivity of the commons
leads to optimal investments that are finite 
even when investment costs vanish, viz.\
when $c\to0$. Later on we will show that
it does not matter whether the size
of the commons is formally infinite, 
as defined (\ref{TOC_E_i}), or finite 
from the start.

\subsection{Selfish individuals}

Selfish agents increase their investments $x_i$ 
until the individual gradients $dE_i/dx_i$ 
vanish. This leads to
\begin{equation}
(1-x_i)\mathrm{e}^{-x_{\rm tot}}=c_i,
\qquad\quad
x_i = 1-c_i\mathrm{e}^{x_{\rm tot}}\,,
\label{TOC_gradient_zero}
\end{equation}
in analogy to the case of an individual
investor discussed above. Here we used that
$dx_{\rm tot}/dx_i=1$. In the stationary
state, individual investments are strictly
linear in the $c_i$. Eq.~(\ref{TOC_gradient_zero})
describes $N$ self-consistency conditions for
$N$ variables, the individual $x_i$, which are
coupled through $x_{\rm tot}$. 
\footnote{In physics terms \cite{gros2015complex,landee2014gentle},
the cumulative investment acts as a 
mediating molecular field.} The size of the
total investment $x_{\rm tot}$ can be determined 
by averaging (\ref{TOC_gradient_zero}) over all agents,
\begin{equation}
\left(1-\frac{x_{\rm tot}}{N}\right)
\mathrm{e}^{-x_{\rm tot}}= \bar{c},
\qquad\quad
\bar{c} = \frac{1}{N}\sum_i c_i\,,
\label{TOC_x_tot_optimal}
\end{equation}
where $\bar{c}$ is the mean per unit cost. The cumulative
investment is hence solely a function of $N$
and of the average costs. In particular it does 
not depend on the actual distribution of the respective
$c_i$. A finite solution exists for all $0<\bar{c}<1$,
as illustrated in Fig.~\ref{fig_x_tot_c_bar}.
In the large-$N$ limit one finds 
$x_{\rm tot}\to \log(1/\bar{c})$.

%
%

\subsection{Initial decimation}

The investment $x_i$ resulting from 
self-centered optimization is
\begin{equation}
x_i = 1-\frac{c_i}{c_{\rm max}},
\qquad\quad
c_{\rm max}=\mathrm{e}^{-x_{\rm tot}}
\label{TOC_c_max}
\end{equation}
which follows from (\ref{TOC_gradient_zero}). Per-unit 
costs exceeding $c_{\rm max}$ lead to $x_i<0$, 
which would result in negative payoffs. When 
this happens, the agent in question is assumed 
to quit the market. Otherwise, agents continue 
to exploit the common resource independently 
of the absolute size of the profit made. It
is assumed that it does not matter for the 
individual whether other agents receive 
smaller or larger payoffs, as long as profits
can be made. In the framework examined here,
direct competition between agents is absent.

The Nash equilibrium depends via (\ref{TOC_x_tot_optimal})
on $\bar{c}$ and $N$, which are both reduced 
when an agent with negative payoffs drops out of the
market. This will lead to a reduced profitability 
limit $c_{\rm max}=\exp(-x_{\rm tot})$, with
the consequence that additional agents may 
become unprofitable.
This iterated process is denoted `decimation'.

For the initial decimation process one starts with 
a given distribution for the per-unit costs $c_i$. One 
then calculates $\bar{c}$ and $x_{\rm tot}$, the 
latter from (\ref{TOC_x_tot_optimal}). Next
$c_{\rm max}$ is determined and agents with 
$c_i>c_{\rm max}$ eliminated. One starts over
again with renormalized quantities:
\begin{equation}
N = \sum_{i,E_i>0} 1,
\qquad\quad
\bar{c} = \frac{1}{N}\sum_{i,E_i>0} c_i\,.
\label{TOC_E_i_positive}
\end{equation}
This procedure is iterated until convergence.
From now on only the population surviving 
the initial decimation process is considered, 
with $N$ and $\bar{c}$ corresponding respectively
to the number and to the average per-unit costs
of the surviving agents.

\begin{figure}[!t]
\centerline{
\includegraphics[width=0.75\columnwidth]{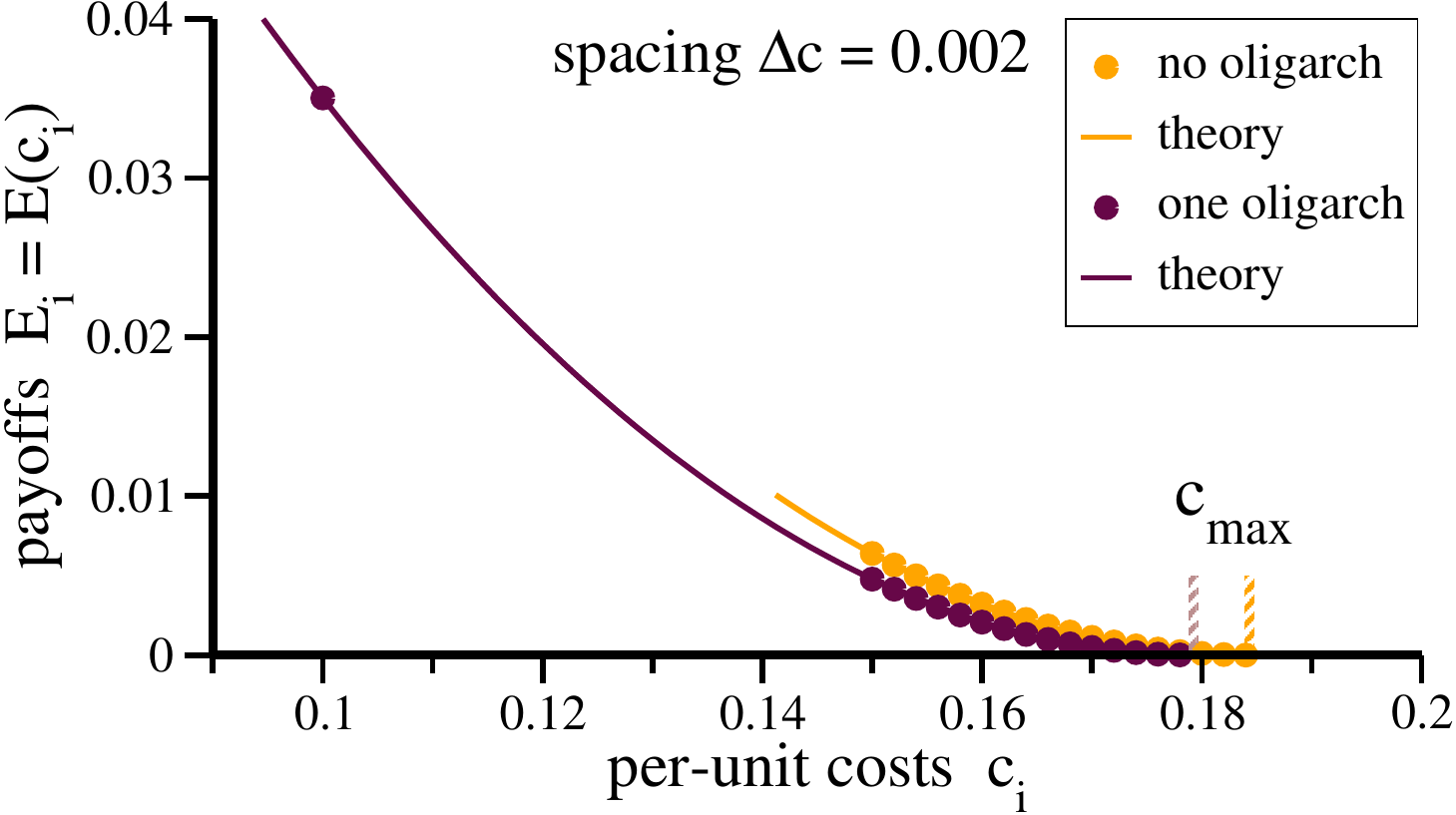}
           }
\caption{{\bf Dispersion relations.}
For a spacing of $\Delta c=0.002$ of consecutive
per-unit costs, the payoffs $E_i$ obtained from
numerical simulations (filled circles). Given
in comparison are the analytic results (lines), 
see (\ref{TOC_dispersionRelation}).
{\bf Orange:} The payoffs for the surviving
$N=18$ agents, with $\bar{c}=0.167$ and
$c_{\rm max} = 0.184$ (striped vertical orange bar). 
Per-unit costs start at $c_{\rm min}=0.15$.
{\bf Brown:} One additional agent, the oligarch,
is added at $c_i=0.1$. One has now 
$\bar{c}=0.16$, $c_{\rm max} = 0.179$ 
(striped vertical brown bar) and $N=16$ total surviving
agents (15 with $c_i\ge0.15$ plus the oligarch).
\label{fig_numerical_00}
}
\end{figure}

\subsection{Dispersion relation}

Combining (\ref{TOC_c_max}) and (\ref{TOC_E_i}) 
results in
\begin{equation}
E_i\big|_{\rm opt} = 
x_i\left(\mathrm{e}^{-x_{\rm tot}}-c_i\right)\Big|_{\rm opt}
=\left(1-\frac{c_i}{c_{\rm max}}\right)
\big(c_{\rm max}-c_i\big)\,,
\label{TOC_E_i_optimal}
\end{equation}
or
\begin{equation}
E(c_i) \equiv E_i\big|_{\rm opt} = c_{\rm max}
\left(1- \frac{c_i}{c_{\rm max}} \right)^2\,.
\label{TOC_dispersionRelation}
\end{equation}
This expression holds for the stationary state,
viz.\ for the Nash equilibrium. It is universal in 
the sense that it depends on $\bar{c}$ and $N$ 
only implicitly via $c_{\rm max}$, but not explicitly.

Eq.~(\ref{TOC_dispersionRelation}) has the
form of a dispersion relation, a notation borrowed
from physics,\footnote{In physics, the dispersion 
relation describes the functional dependency of 
the frequency (or energy) of an excitation
with respect to the controlling variable, in 
general the wavevector \cite{colombo2021solid}. 
In analogy, Eq.~(\ref{TOC_dispersionRelation}) 
specifies the relation of the payoff $E_i$ 
(taken to correspond to the energy) to 
the controlling degree of freedom, in this
case the per-unit cost $c_i$.
} providing a distinct functional relation 
between $c_i$ and the respective payoff.
It is quite remarkable that the dispersion 
is strictly quadratic over the entire
range, $c_i\in[0,c_{\rm max}]$, and not only 
close to marginal profitability. In the latter 
regime, viz.\ when $c_i$ is close to $c_{\rm max}$,
the quadratic dependency arises because
the optimal investment $x_i=1-c_i/c_{\rm max}$ 
vanishes linearly when $c_i\to c_{\rm max}$.


\subsection{Catastrophic poverty}

It follows from (\ref{TOC_x_tot_optimal}),
that
\begin{equation}
\frac{\bar{c}}{c_{\rm max}} = 
\frac{\bar{c}}{\mathrm{e}^{-x_{\rm tot}}}
= 1-\frac{x_{\rm tot}}{N}
\approx 1-\frac{\log(1/\bar{c})}{N}\,,
\label{TOC_c_bar_max}
\end{equation}
where we used in the last step the large-$N$
expression $x_{\rm tot}\to\log(1/\bar{c})$
for the overall investment. The relative
distance of the profitability limit $c_{\rm max}$
to the mean per-unit costs $\bar{c}$ vanishes
hence in the large $N$ limit when $\bar{c}>0$.
This has dramatic consequences for the size
of the payoffs of less competitive investors.

Inserting (\ref{TOC_c_bar_max}) into the dispersion 
relation,
$E_i|_{\rm opt}  = c_{\rm max}(1-c_i/c_{\rm max})^2$,
yields
\begin{equation}
E_i\big|_{\rm opt} \
\stackrel{c_i=\bar{c}}{\longrightarrow} \
\frac{\bar{c}}{1-x_{\rm tot}/N}
\left(\frac{x_{\rm tot}}{N}\right)^2
\sim \frac{1}{N^2}
\label{TOC_E_i_optimal_bar_c}
\end{equation}
for the payoff of an average agents,
defined by $c_i=\bar{c}$. We assumed 
for the last step that $x_{\rm tot}$
is not diverging with $N$, which is
the case when one considers a sequence
of populations with increasing sizes,
but fixed average per-unit costs 
$\bar{c}$. The result is, somewhat
surprisingly, that the optimal 
individual payoff scales with $(1/N)^2$. 

Eq.~(\ref{TOC_E_i_optimal_bar_c}) is a 
non-trivial result. A finite cumulative 
payoff $E_{\rm tot} = \sum_i E_i$ can be 
extracted from the common resource and 
one could have expected that everybody gets
a share scaling with $1/N$, with the actual
size being a function of individual per-unit costs 
$c_i$. This is however not the case. The 
balance between costs and returns becomes
increasingly fine-tuned with larger and large
$N$, which drives $c_{\rm max}$ successively 
closer to $\bar{c}$.

Payoffs drop continuously with increasing
per-unit costs, $c_i$, which means that
the $(1/N)^2$ scaling following from
(\ref{TOC_E_i_optimal_bar_c}) holds at least
for all $\bar{c}\le c_i<c_{\rm max}$. In
general there is a finite fraction of agents in 
the interval $\bar{c} \le c_i < c_{\rm max}$,
typically about half the population is
situated in this region. The payoffs of 
these agents are not just small, but 
functionally depressed, in the sense that
payoff reduction is due to a specific functional
dependency on $N$. This situation is 
denoted here as `catastrophic poverty'.

The origins of catastrophic poverty can be
traced to three properties of the dispersion
relation $E(c_i)$: 
\begin{itemize}
\item $c_{\rm max}$. An upper bound for per-unit
costs exists. Investments become non-profitable 
above, with $E(c_{\rm max})=0$.
\item $c_{\rm max}-\bar{c}\sim 1/N$. The mean
per-unit costs approach the upper bound when the
number of agents increases. Compare (\ref{TOC_c_bar_max}).
\item $E(c_i) \sim (c_{\rm max}-c_i)^2$. A quadratic
dispersion relation implies that payoffs vanish 
quadratically close to $c_{\rm max}$.
\end{itemize}
Any Nash state with a dispersion relation 
fulfilling these three properties is 
characterized by catastrophic poverty. We
will show later on that this is the case
not only for the basic model specified by
Eq.~(\ref{TOC_E_i}), but for a wide class
of generic frameworks describing the 
exploitation of a common resource by 
selfish agents.

\subsection{Wealthy oligarchs}

Catastrophic poverty affects a substantial
part of the population, but not necessarily 
all agents. A few individuals, the oligarchs, 
may have substantial payoffs. Whether
oligarchs exist depends on the 
distribution $\{c_i\}$ of the 
per-unit costs of the surviving agents.

We start with two illustrative examples 
for the cost-distribution $\{c_i\}$. In 
the first case, the per-unit costs of the 
$N$-agents are uniformly distributed
around the mean, in the interval 
$[\bar{c}-\delta c, \bar{c}+\delta c]$, 
where $\delta c < c_{\rm max}-\bar{c}$.
The distance to the boundary, $c_{\rm max}-c_i$,
scales therefore as $1/N$ for all agents. Given
that the dispersion relation $E(c_i)$ is
quadratic, one has $E(c_i)\sim (1/N)^2$ 
for all agents. In this example, catastrophic 
poverty affects the entire population.

In the second example a single agent has
vanishing production costs, $c_i=0$, with
the other agents having the identical
$c_i$, located at $\bar{c}+\delta c$,
with 
$\delta c = \alpha (c_{\rm max}-\bar{c})$
and $\alpha <1$.
The average is reproduced for
\begin{equation}
\bar{c} = \frac{N-1}{N}\big[\bar{c}+\alpha (c_{\rm max}-\bar{c})\big],
\qquad\quad
\alpha = \frac{1}{N-1}
\label{oligarch_bar_c}\,\frac{\bar{c}}{c_{\rm max}-\bar{c}}\,.
\end{equation}
Using (\ref{TOC_c_bar_max}) leads to
\begin{equation}
\alpha = \frac{N}{N-1}\,\frac{1}{x_{\rm tot}}
\approx \frac{1}{x_{\rm tot}}\,.
\label{oligarch_bar_c_condition}
\end{equation}
The profitability condition $\alpha<1$ is
satisfied for $x_{\rm tot}>1$, which holds
for a large range of $\bar{c}$ and $N$, 
as evident from the data presented 
Fig.~\ref{fig_x_tot_c_bar}. 
The payoff of the oligarch is
\begin{equation}
E(0) = c_{\rm max} = \frac{\bar{c}}{1-x_{\rm tot}/N}\,,
\label{oligarch_E_0}
\end{equation}
which is finite whenever $\bar{c}>0$. 
Note that (\ref{TOC_c_bar_max}) has been used
for $c_{\rm max}$ and that (\ref{oligarch_E_0})
holds for $N$ that are large enough for
(\ref{oligarch_bar_c_condition}) to be fulfilled.
As a consequence, the profit made by the oligarch
can be orders of magnitudes larger than the 
profit $E(c_i)=(c_{\rm max}-c_i)^2/c_{\rm max}$
of any of the remainder $N\!-\!1$ agents, 
\begin{equation}
E(c_i) = \frac{(x_{\rm tot}-1)^2}{x_{\rm tot}^2}\,
\frac{(c_{\rm max}-\bar{c})^2}{c_{\rm max}},
\qquad\quad
c_i=\bar{c}+\frac{c_{\rm max}-\bar{c}}{x_{\rm tot}}\,,
\label{oligarch_E_N-1}
\end{equation}
which are all suffering from catastrophic poverty
when the size of the population is substantial.

When oligarchs have finite payoffs, as in
the above example, the cumulative payoff 
is also finite. For the general case we
reorganize the dispersion relation
(\ref{TOC_dispersionRelation}) as
\begin{equation}
E_i(c_i) = 
c_{\rm max} - 2c_i +
\frac{c_i^2-\bar{c}^2+\bar{c}^2}{c_{\rm max}}\,,
\label{oligarch_E_plusMinus}
\end{equation}
where we added and subtracted $\bar{c}^2$ 
in the numerator of the last term. Taking the 
average over all agents yields
$\bar{E}= \sum_i E_i(c_i)/N$,
\begin{equation}
\bar{E} = \frac{(c_{\rm max}-\bar{c})^2}{c_{\rm max}}
+\frac{\sigma_c^2}{c_{\rm max}}\,,
\label{oligarch_E_bar}
\end{equation}
where $\sigma_c^2=\sum_i(c_i^2-\bar{c}^2)/N$
is the variance of the per-unit costs.
The first term in (\ref{oligarch_E_bar})
scales as $(1/N)^2$, as shown previously,
see (\ref{TOC_c_bar_max}). The overall scaling 
of $\bar{E}$ depends therefore on the
scaling of the variance $\sigma_c^2$ with $N$.

For the first example discussed above we have
$\sigma_c\sim1/N$. Consistently, the average 
payoff is proportional to $(1/N)^2$. The 
cumulative payoff $N\bar{E}$ scales
therefore as $1/N$, vanishing in the limit
of large populations. Everybody suffers.

In our second example an oligarch with a finite 
payoff is present, which implies a finite cumulative
payoff $N\bar{E}$ and that $\bar{E}$ scales as 
$1/N$. This is consistent with the scaling of the 
variance $\sigma_c^2$, to which the per-unit cost 
$c_i=0$ of the oligarch contributes a term 
$\bar{c}^2/N$. The upshot is, that the presence 
of one or more oligarchs increases average payoffs, 
however without eliminating catastrophic poverty
for the majority of agents.

Conceptually, our approach is based on the analysis
of a series of distributions $\{c_i\}$ for the 
player specific per-unit costs, with the premise that these
distributions can be defined consistently for various 
population sizes $N$. In practice, the final distribution 
$\{c_i\}$ of the surviving agents results from
the initial decimation process. Performing numerical 
simulations, as described further below in more
detail, we found that the two types of Nash-states
discussed here appear readily from decimation
for a wide range of starting allocations for
the agent-specific cost structure.

\subsection{Cooperation}

For the analysis performed so far we assumed  
that agents do not cooperate. For comparison 
we investigate now the possible benefits of
coordinating investments. Our purpose is however
not to develop a generic theory of cooperation,
for which there would be a range of distinct 
optimization principles. Cooperation is 
uniquely defined only when agents are identical,
viz.\ when $c_i\equiv c$. In general, an objective 
could be to optimize the mean payoff 
$\bar{E}$, either alone or in conjunction
with a given criterion for the fairness of income 
distributions, like the Gini index. An 
additional variable is the number of surviving 
agents. It could be desirable to keep the size 
of the population as large as possible when 
investing cooperatively, or not. Here we 
restrict ourselves to a basic discussion.

As a reference protocol we consider that agents 
invest equal shares, $x_i\equiv x_{\rm tot}/N$, 
which is optimal for the case of identical agents.
The aim is then to optimize total payoff, 
$E_{\rm tot} =\sum_i E_i$, namely
\begin{equation}
E_{\rm tot} = (c_{\rm max}-\bar{c})x_{\rm tot},
\qquad\quad
c_{\rm max} = \mathrm{e}^{-x_{\rm tot}}\,.
\label{TOC_E_tot_coop}
\end{equation}
In effect, the community of agents acts as a 
single investor, see (\ref{TOC_E_i_one}) 
and (\ref{TOC_E_tot_one_optimal}). The consequence
is that the cumulative payoff is finite when 
agents cooperate. For the individual agent 
we have
\begin{equation}
E_i\big|_{\rm coop} = \big(c_{\rm max}-c_i\big) 
\frac{x_{\rm tot}}{N}\,,
\label{TOC_E_i_coop}
\end{equation}
which scales as $1/N$. Catastrophic poverty
is avoided when cooperating, with each surviving
agent receiving a fair share. This is consistent
with the result that the dispersion relation is 
linear, as expressed by (\ref{TOC_E_i_coop}),
which corresponds to the classical expectation. 
The overall level of exploitation remains
comparatively low, as $x_{\rm tot}<1$ for 
$N=1$, as shown in Fig.~\ref{fig_x_tot_c_bar}. 
If follows from $c_{\rm max}= \exp(-x_{\rm tot})$ 
that $c_{\rm max}<1$. High-cost agents will
therefore be driven out of the market 
during the initial decimation process even 
when everybody cooperates according to
the here considered protocol. In contrast,
decimation will have no effect when agents
have identical investment costs.

\subsection{Generic productivity function}

Catastrophic poverty arises in large classes 
of models describing the exploitation
of a common resource by non-cooperating
agents. In the basic model (\ref{TOC_E_i}),
the commons degraded exponentially with total 
investments, as $\exp(-x_{\rm tot})$.
On a general level we denote with $P(x_{\rm tot})$ 
the productivity of the commons. The 
respective payoff function is then
\begin{equation}
E_i = x_i\big[P(x_{\rm tot})-c_i\big],
\qquad\quad
P(0)=1,
\qquad\quad
P'(x_{\rm tot})<0\,.
\label{P_E_i}
\end{equation}
Payoffs are optimal for
\begin{equation}
P(x_{\rm tot})-c_i = -x_iP'(x_{\rm tot}),
\qquad\quad
E_i\big|_{\rm opt} = -P'(x_{\rm tot})\,x_i^2\,,
\label{P_E_i_optimal}
\end{equation}
with $x_{\rm tot}$ being determined by averaging the
first equation over all agents,
\begin{equation}
P(x_{\rm tot})-\bar{c} = -\frac{x_{\rm tot}}{N}P'(x_{\rm tot})\,.
\label{P_x_optimal}
\end{equation}
The maximal per-unit costs $c_{\rm max}$ 
follows from the limit $x_i\to0$ in (\ref{P_E_i_optimal}),
\begin{equation}
c_{\rm max} = P(x_{\rm tot}),
\qquad\quad
x_i = \frac{c_{\rm max}-c_i}{-P'(x_{\rm tot})}\,,
\label{P_c_max}
\end{equation}
which leads to
\begin{equation}
E_i\big|_{\rm opt} = 
\frac{(c_{\rm max}-c_i)^2}{-P'(x_{\rm tot})}\,.
\label{P_disperisonRelation}
\end{equation}
The dispersion relation is again strictly 
quadratic. It now depends explicitly on 
$x_{\rm tot}$, in contrast to the original 
case, see (\ref{TOC_dispersionRelation}), 
which is however a minor alternation.
Furthermore it follows from (\ref{P_x_optimal})
that $c_{\rm max}-\bar{c}\sim 1/N$, the 
second precondition for catastrophic poverty.
Catastrophic poverty arises hence for all monotonically
decaying  productivity functions $P(x_{\rm tot})$.

In (\ref{P_E_i}) the individual investment costs 
are strictly linear in $x_i$. Further 
on we will relax this assumption and discuss 
in detail the extension of the basic model to 
convex and concave cost functions. Another possible 
extension of (\ref{TOC_E_i}) regards the value
a certain amount of extracted resources has
for the individual investor. This value may 
differ by a factor $r_i$ between agents.
The corresponding model,
\begin{equation}
E_i = x_i\big(r_i P(x_{\rm tot})-c_i\big) = 
x_ir_i\left(P(x_{\rm tot})-\frac{c_i}{r_i}\right)\,,
\label{P_r_i}
\end{equation}
is functionally equivalent to (\ref{P_E_i}). 
It then follows that the dispersion relation, 
which can be derived now for $\widetilde{E}_i=E_i/r_i$, 
is still quadratic, albeit not as a function 
of $c_i$, but of $\tilde{c}=c_i/r_i$.

\subsection{Runaway exploitation}

Our analysis is based on the condition that
the level of the total investment does not
diverge in the Nash equilibrium. This is indeed 
the case whenever $\bar{c}$ and $N$ are finite,
namely when $\bar{c}>0$ and $N<\infty$. For
the original productivity function,
$P=\exp(-x_{\rm tot})$, total investment remains
finite even in the limit $\bar{c}\to0$, for which 
we found from (\ref{TOC_x_tot_optimal}) that 
$x_{\rm tot}\to N$. It can however happen,
for other $P(x_{\rm tot})$, that the optimality
condition (\ref{P_x_optimal}) has no solution
when $\bar{c}\to0$. As an example we consider 
the productivity function
\begin{equation}
P(x_{\rm tot}) = \frac{1}{(1+x_{\rm tot})^\beta}\,,
\label{run_P_x}
\end{equation}
which decays as a power-law for large
total investments. The respective payoff
function has a well-defined maximum for $N=1$
and $\bar{c}=0$ when $\beta>1$. For general values
of $N$, the optimality condition (\ref{P_x_optimal}) 
reduces to
\begin{equation}
x_{\rm tot}\big|_{\rm opt} \
\stackrel{\bar{c}\to0}{=} \ \frac{N}{\beta-N}
\label{run_condition}
\end{equation}
when $\bar{c}\to0$. The optimal total investment 
diverges hence even for finite of numbers of
agents, namely whenever $N\ge\beta$. Instead, 
taking first the limit $N\to\infty$ leads to
$x_{\rm tot} = (1/\bar{c})^{1/\beta}-1$, 
which is positive and finite for all
$0<\bar{c}<1$. Runaway exploitation can hence
occur for finite $N$, but only when average
per-unit costs are formally zero\footnote{In 
the real world, costs can be small, but they 
are never exactly zero. The limit $\bar{c}\to0$
corresponds in this sense to a formal consideration.}

One could characterize a given environment
as resilient when its productivity decays
only slowly with increased exploitation
efforts. An example would be a fishing ground
with robust regrowth rates. It is interesting
in this regard that exponential functions
decay faster with increasing argument than a 
power law. The possible occurrence of runaway 
exploitation for power-law productivity functions 
suggests hence the hypothesis that the likelihood 
that existing environments will suffer from 
over-exploitation may raise with resilience.
This conclusion is consistent with the results
from a study of ecosystem exploitation that
includes corruption traits \cite{lee2017games}.

\subsection{Finite-size commons}

Hitherto, investments of any size would 
result in positive nominal returns. 
Alternatively one can consider a 
common-pool resource for which there is 
an upper bound for total investments, 
$x_{\rm max}$. At this point, the productivity 
of the commons vanishes altogether. For 
simplicity we denote $x_{\rm max}$ as the 
size of the commons \cite{ostrom1999coping}.
A possible productivity function is
\begin{equation}
P(x_{\rm tot}) = 1-\frac{x_{\rm tot}}{x_{\rm max}}\,,
\label{ostrom_linear}
\end{equation}
which leads via (\ref{P_x_optimal}) to the
optimal total investment
\begin{equation}
x_{\rm tot}\big|_{\rm otp} = 
(1-\bar{c})\,\frac{N}{N+1}\, x_{\rm max}\,.
\label{ostrom_x_tot}
\end{equation}
This expression is functionally well behaved
in the sense that runaway exploitation is absent
for all parameter regimes, in contrast to
(\ref{run_condition}). The optimal
overall investment approaches the carrying
capacity $x_{\rm max}$ of the commons
when  $\bar{c}\to0$ and $N\to0$. The dispersion 
relation (\ref{P_disperisonRelation}) 
holds with $P'(x_{\rm tot})=-1/x_{\rm max}$,
which implies that catastrophic poverty
is present. Catastrophic poverty will hence
emerge for both finite-size and formally 
unbounded commons. We note that
(\ref{ostrom_linear}) was used by Ostrom 
for a laboratory protocol 
\cite{ostrom1999coping}, as discussed 
later on in further detail.

\section{Simulations}

We solved the self-consistency condition 
(\ref{TOC_x_tot_optimal}) for $x_{\rm tot}$
numerically, which allows to determine
$c_{\rm max} = \exp(-x_{\rm tot})$ and
individual investments $x_i$, the latter 
from the linear relation (\ref{TOC_gradient_zero}). 
As expected, one finds perfect agreement
between the direct evaluation of the 
payoffs, via $E_i=(c_{\rm max}-c_i)x_i$,
and the prediction of the dispersion 
relation (\ref{TOC_dispersionRelation}).

Numerically accessible is in particular the
initial decimation procedure. Starting with 
a given distribution $\{c_i\}$ for the 
per-unit costs, one performs the decimation 
process iteratively until the number of surviving 
agents does not change any more. We concentrate 
here on distributions $\{c_i\}$ for which the bulk
of the per-unit costs are uniformly spaced,
$c_i=c_{\rm min}+i\Delta c$, compare
Fig.~\ref{fig_numerical_00}. To the set
of agents with equally spaced per-unit costs 
we added at most an additional agent, the
oligarch, with a per-unit cost well 
below $c_{\rm min}$. 

For the simulation presented in
Fig.~\ref{fig_numerical_00} we
started with $N_{\rm start}$ agents.
Without an oligarch, $N=18$ agents
remain after decimation when using 
$c_{\rm min}=0.15$ and $\Delta c=0.002$.
The exact number of starting agents is 
irrelevant, as long as $N_{\rm start}\ge18$,
as agents with $c_i>c_{\rm max}$ are decimated 
out in any case. Lower numbers of $N_{\rm start}$ 
would lead in contrast to a different Nash
state. For example, no decimation is
performed when starting with a single
agent.

\begin{table}[!b]
\caption{{\bf Simulation results.}
For regularly spaced per-unit costs, with
$c_{\rm min}=0.15$ and $\Delta c = 0.002$,
the properties of the stationary state resulting
from the initial decimation procedure. Compare
Fig.~\ref{fig_numerical_00}. Given is the number
$N$ of surviving agents, total investment and
payoff, $x_{\rm tot}$ and $E_{\rm tot}$, together
with the average and maximal per-unit
costs $\bar{c}$ and $c_{\rm max}$. The last two
columns indicate if an oligarch at $c_i=0.1$ 
was included and whether agents did cooperate.
}
\medskip
\centerline{\framebox{
\begin{tabular}{r|llll|ll}
N & $x_{\rm tot}$ & $E_{\rm tot}$ & $\bar{c}$ & 
$c_{\rm max}$& oligarch & coop.\ \\ \hline 
18   & 1.691 & 0.040 & 0.167 & 0.184 & no & no\\
1    & 0.698 & 0.243 & 0.150 & 0.497 & no & no\\
18   & 0.673 & 0.231 & 0.167 & 0.510 & no & yes\\ \hline
15+1 & 1.720 & 0.061 & 0.160 & 0.179 & yes & no\\
 1+1 & 1.184 & 0.219 & 0.125 & 0.306 & yes & no \\ 
15+1 & 0.683 & 0.236 & 0.160 & 0.505 & yes & yes
\end{tabular}
}}
\label{table_numerics}
\end{table}

\subsection{No oligarchs}

We first examine the results when there
is no oligarch present. All calculations 
are for the reference model, as defined by
(\ref{TOC_E_i}), and for two final configurations, 
$N=18$ and $N=1$ (obtained by starting respectively 
with $N_{\rm start}\ge18$ and $N_{\rm start}=1$).
The data is given in Table~\ref{table_numerics}.
We recall that the productivity of the commons,
\begin{equation}
P = \mathrm{e}^{-x_{\rm tot}} = c_{\rm max},
\label{NUM_d_c}
\end{equation}
coincides with the maximum per-unit cost,
which is given in Table~\ref{table_numerics}.
Alternative one can regard $P=c_{\rm max}$
as a measure for the degree of degradation, 
viz.\ for the status of the common resource.

Comparing the results for $N=18$ and $N=1$
one notices that the respective $\bar{c}$ differ 
by only about 10\%. The total payoff $E_{\rm tot}$
is however reduced by a factor $6\approx0.243/0.04$ 
when increasing the population from $N=1$ to
$N=18$. This reduction can be seen as a 
precursor of catastrophic poverty, as 
expressed by Eq.~(\ref{TOC_E_i_optimal_bar_c}).
For a given $c_{\rm min}$ the number of 
surviving agents depends on the spacing 
$\Delta c$, with $N$ increasing when
$\Delta c$ is decreased. We tested that the
$(1/N)^2$ scaling characterizing catastrophic
poverty on an individual level does indeed 
hold when $N$ is further increased.

The numerical results show that overall welfare 
suffers substantially already for comparatively 
modest numbers of independently exploiting actors, 
here $N=18$. An increased population size
leads also to a deteriorating state of the 
commons, but to a somewhat lesser extend.
The ratio of the productivity function 
$P=c_{\rm max}$ for $N=1$, and respectively
for $N=18$, is $0.497/0.184=2.7$, 
compare (\ref{NUM_d_c}), which is about half 
the respective payoff ratio, $0.243/0.04=6.1$.
This effect will become less important for further 
increased population sizes, as $c_{\rm max}$ is 
bounded from below by $\bar{c}$.

Also included in Table~\ref{table_numerics}
is the data for $N=18$ cooperating agents,
using the basic protocol discussed further 
above, which is defined by $x_i\equiv x_{\rm tot}/N$.
By construction, the mean $\bar{c}$ does
not depend on whether agents do or do not 
cooperate.  When cooperating, $E_{\rm tot}$ 
is substantially higher with respect to the
same number on non-cooperating agents. The 
total payoff of cooperating agents is identical
to that of a single agent with $c_i=\bar{c}$,
but a bit reduced with respect to the case
$N=1$ shown in Table~\ref{table_numerics},
which is for a single agent with $c_i=c_{\rm min}$.
In any case, cooperating agents do not suffer 
from catastrophic poverty.

\subsection{A single oligarch}

We added an additional agent with $c_i=0.1$
to the setup described before. Without 
cooperation $N=16=15+1$ agents survive 
decimation, the oligarch plus 15 bulk 
investors with $c_i\ge c_{\rm min}=0.15$.
The data is included in Table~\ref{table_numerics}
and shown in Fig.~\ref{fig_numerical_00}. 
Nominally, the total payoff is larger 
when an oligarch is present. Calculating 
the profits of the 15 bulk agents one
finds a cumulative payoff of $0.0267$,
which is less than the payoff of the oligarch,
$0.035$. Overall, the addition of a single 
oligarch has only an indirect influence on
the welfare of the remainder agents. 

\begin{figure}[!t]
\centerline{
\includegraphics[width=0.75\columnwidth]{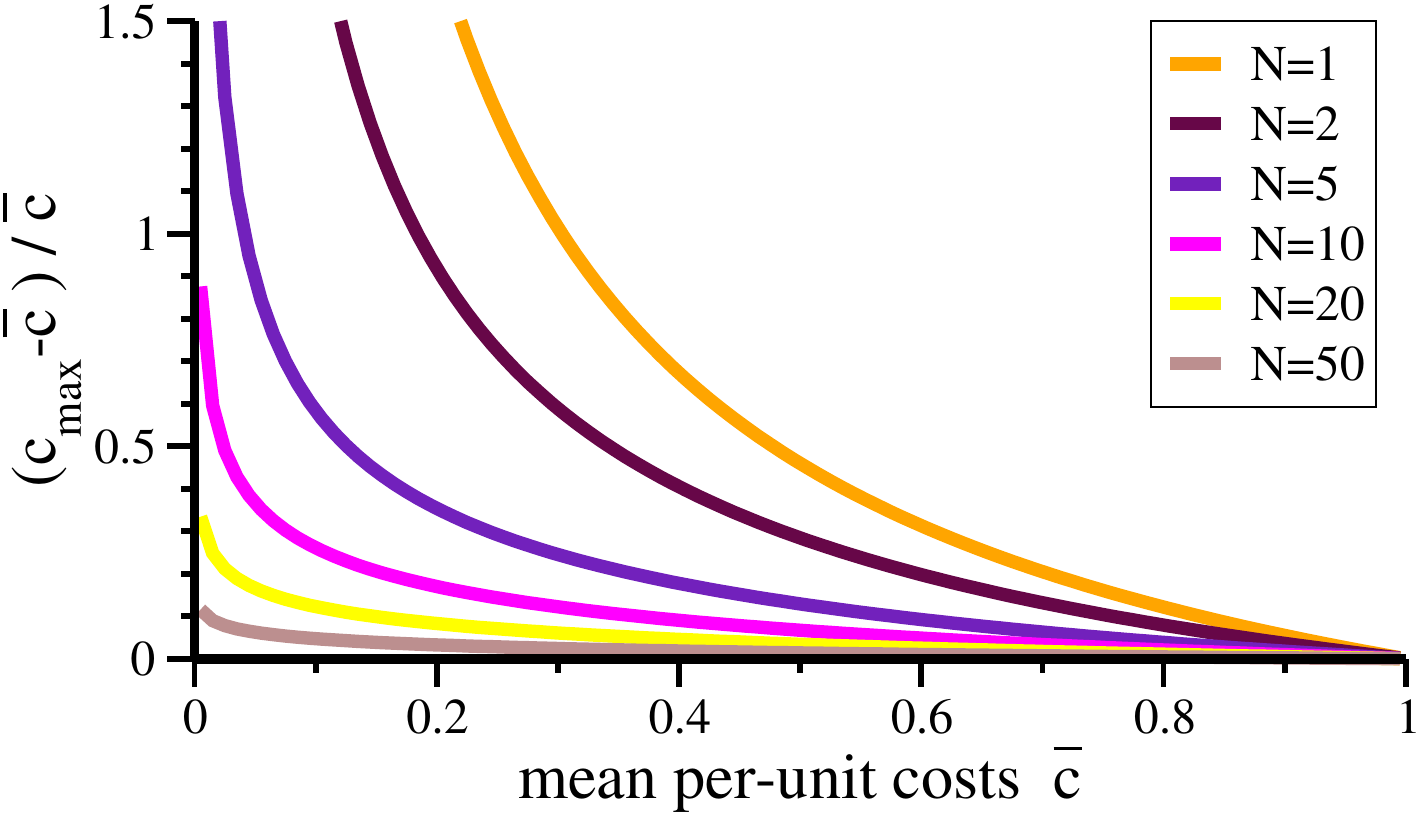}
           }
\caption{{\bf Window of participation.}
As determined by (\ref{TOC_c_max}), the relative
distance between maximal and mean per-unit costs, 
$\delta c_{\rm p}=(c_{\rm max}-\bar{c})/\bar{c}$. 
The data is for fixed numbers $N$ of agents. E.g., 
adding  an agent to the case $N=1$ will shift the 
curve to $N=2$. The large-$N$ limiting behavior is
$\log(1/\bar{c})/N$. For fixed $N$, the data
diverges for vanishing mean per-unit costs.
The window of participation for less efficient 
agents, $\delta c_{\rm p}$, can be interpreted 
also in term of relative profit margins, 
see Eq.~(\ref{NUM_M_c_i}).
\label{fig_relative_c_max}
}
\end{figure}

\subsection{Participation window}

The per-unit cost of less efficient 
agents is located typically in the 
interval $[\bar{c},c_{\rm max}]$, 
which can hence be regarded to represent 
a window of participation. It is of 
interest to investigate how large this 
region is in relative terms. 
In Fig.~\ref{fig_relative_c_max} 
the results for
\begin{equation}
\delta c_{\rm p} = \frac{c_{\rm max}-\bar{c}}{\bar{c}} 
= \frac{x_{\rm tot}}{N-x_{\rm tot}}\,,
\label{NUM_delta_c_max}
\end{equation}
are shown, where we used (\ref{TOC_c_bar_max}) 
for $c_{\rm max}$. Note that 
$\lim_{\bar{c}\to0} x_{\rm tot} = N$
holds, as shown in Fig.~\ref{fig_x_tot_c_bar},
which implies that the relative participation 
window $\delta c_{\rm p}$ diverges 
for $\bar{c}\to0$. In the large-$N$
limit one has $x_{\rm tot}\to\log(1/\bar{c})$
for $\bar{c}>0$ and hence $\delta c_{\rm p}\to\log(1/\bar{c})/N$
when costs are small but finite. 

For fixed population sizes $N$ the curves for 
$\delta c_{\rm p}$ presented in Fig.~\ref{fig_x_tot_c_bar} 
increase monotonically with decreasing average investment
costs. Interestingly, one can regard $\bar{c}$ as
a measure for the mean technological level of the
investing group of agents. In this view, small 
average costs correspond to a technological 
advanced society, which is characterized by 
a large total investment $x_{\rm tot}$ and a 
correspondingly degraded commons. In addition,
technologically advanced societies offer a
substantial window of participation for less 
efficient agents, as expressed by a comparatively large
$\delta c_{\rm p}=(c_{\rm max}-\bar{c})/\bar{c}$,
which implies that a substantial range of 
agents may survive the decimation process.
This implies vice-versa that catastrophic
poverty is less likely to arise in
societies with large average costs $\bar{c}$.

The data for the window of participation
presented in Fig.~\ref{fig_relative_c_max}
can be interpreted in terms of relative
profit margins $M(c_i)$. In the Nash state,
the payoff per investment is
\begin{equation}
M(c_i) = \frac{E_i}{c_i x_i} = \frac{c_{\rm max}-c_i}{c_i}\,,
\label{NUM_M_c_i}
\end{equation}
which follows directly from $E_i= x_i(c_{\rm max}-c_i)$.
For a typical investor, with $c_i=\bar{c}$,
the relative profit margin $M(\bar{c})$ 
coincides with $\delta c_{\rm p}$, 
see Eq.~(\ref{NUM_delta_c_max}). For example, 
for $N=50$ one has $\delta c_{\rm p} \approx 0.12$
for small $\bar{c}$, which corresponds to
a typical profit margin of about 12\%, still 
a healthy value. In the opposite limit, for 
small $N$, profit margins readily surpass 100\% 
for efficient investors.

These considerations suggest two conclusions.
Firstly, that small populations will strongly
attract additional investors in technologically
advanced societies, namely when averages costs
are small. Secondly, efficient investors will
make healthy profits even when $N$ is substantial,
say $N=50$ or larger. At first sight, this
observation seems to rule out the occurrence
of catastrophic poverty. This is however not the
case, as both $c_{\rm max}-\bar{c}$ and $x_i$
scale with $1/N$ for per-unit costs within the 
window of participation, or close to $\bar{c}$
in general.

\section{Convex and concave cost functions}

The exploitation of a common resource
by a group of self-centered agents
can be described by a large class
of models. Our discussions centered so 
far mostly on a specific implementation,
the basic model (\ref{TOC_E_i}), which 
we extended with regard to general 
productivity functions $P(x_{\rm tot})$,
see (\ref{P_E_i}). We examine now the
effect of non-linear cost functions,
\begin{equation}
E_i= x_iP(x_{\rm tot}) - c_iC_i(x_i),
\qquad\quad
C_i(0)=0,
\qquad\quad
C_i'(0)=1\,,
\label{PC_E_i}
\end{equation}
where we did write the cumulative individual
investment costs as $c_iC_i(x_i)$. The parameterization
used reduces to $c_iC_i(x_i)=c_ix_i+O(x_i^2)$
for small investments $x_i$, with the prefactor 
$c_i$ corresponding to the initial marginal costs
\cite{delis2014estimation}.
The occurrence of catastrophic poverty depends
on the functional form of the dispersion relation
close to the profitability limit, viz.\ what
happens when the $x_i$ are small, which is
governed in turn by the value of $c_i$. On
a functional level, $C_i(x_i)$ and $c_i$ correspond
to a dimensionless cost function and the
respective monetary scale.

\subsection{Catastrophic poverty vs.\ entry barriers}

Catastrophic poverty will be present when the
cost function is weakly concave, but not
for strongly concave cost functions. This
can be seen by considering the low-$x_i$
expansion of the dimensionless cost function,
\begin{equation}
C_i(x_i)=x_i-\gamma_i x_i^2+O(x_i^3),
\qquad\quad
\gamma_i = -C_i''/2\,.
\label{PC_c_expansion}
\end{equation}
The dimensionless marginal costs decrease
as $C_i'\approx 1-2\gamma x_i$ when $\gamma_i>0$,
which can lead to the well-known phenomenon
of an entry barrier 
\cite{bain1954economies,schmalensee1981economies}.
Potentially, when economies of scale are
impactful, agents make profits for large, 
but not for small investments. It is then 
not viable for a new actor to enter the 
existing market configuration with an 
initial small investment. When and how
this happens depends in addition on
the productivity function of the commons.
For a detailed discussion, in particular
also in regard to the emergence of catastrophic
poverty, we will examine a concrete model.

\subsection{Concave cost functions}

For concreteness we use 
\begin{equation}
C_i(x_i) = \frac{1}{\gamma_i}\log(1+\gamma_i x_i),
\qquad\quad
C_i'(x_i) = \frac{1}{1+\gamma_i x_i}
\label{CC_P_C}
\end{equation}
for the dimensionless cost function. For $\gamma_i\to0$ 
the original model (\ref{TOC_E_i}) with constant
marginal costs is recovered. To order $x_i^2$, 
the Taylor series of (\ref{CC_P_C}) reduces to 
(\ref{PC_c_expansion}). In general the cost function
specified by (\ref{CC_P_C}) is convex/concave 
respectively for $\gamma<0$ and $\gamma>0$. 
Note that costs diverge logarithmically for
$x_i\to1/|\gamma|$ when $\gamma<0$.

For a reference model with non-linear costs
we set $\gamma_i\equiv\gamma$ and 
$P(x_{\rm tot})=\exp(-x_{\rm tot})$. The payoff 
functional is then
\begin{equation}
E_i = x_i\mathrm{e}^{-x_{\rm tot}}-\frac{c_i}{\gamma}
\log(1+\gamma x_i)\,,
\label{CC_E_i}
\end{equation}
which leads to the gradient
\begin{equation}
\frac{dE_i}{dx_i} = 
(1-x_i)\mathrm{e}^{-x_{\rm tot}} - \frac{c_i}{1+\gamma x_i}\,.
\label{CC_gradient}
\end{equation}
The locus of the $x_i=0$ crossing of the 
gradient $E_i'=dE_i/dx_i$ is given by 
$\exp(-x_{\rm tot})=c_{\rm max}$. It holds 
as before that $c_{\rm max}$ coincides 
with the profitability boundary for moderate 
values of $\gamma$. The situation changes 
however for larger values of $\gamma$, as 
shown further below. Setting $E_i'=0$ yields
\begin{equation}
1 + x_i(\gamma-1) - \gamma x_i^2 = c_i\mathrm{e}^{x_{\rm tot}},
\qquad\quad
x_i = \frac{\gamma-1}{2\gamma}\pm
\sqrt{\left(\frac{\gamma-1}{2\gamma}\right)^2
+\frac{c_{\rm max}-c_i}{\gamma\,c_{\rm max}} }\,.
\label{CC_x_i_roots}
\end{equation}
The negative/positive root corresponds to the 
dynamical stable solution respectively for 
$\gamma<0$ and $\gamma>0$, which can be shown 
by a standard stability analysis. In both cases 
the $\gamma=0$ result 
$x_i=(c_{\rm max}-c_i)/c_{\rm max}$ is recovered
in the limit $\gamma\to0$. Compare (\ref{TOC_gradient_zero}).

\begin{figure}[!t]
\centerline{
\includegraphics[width=0.75\columnwidth]{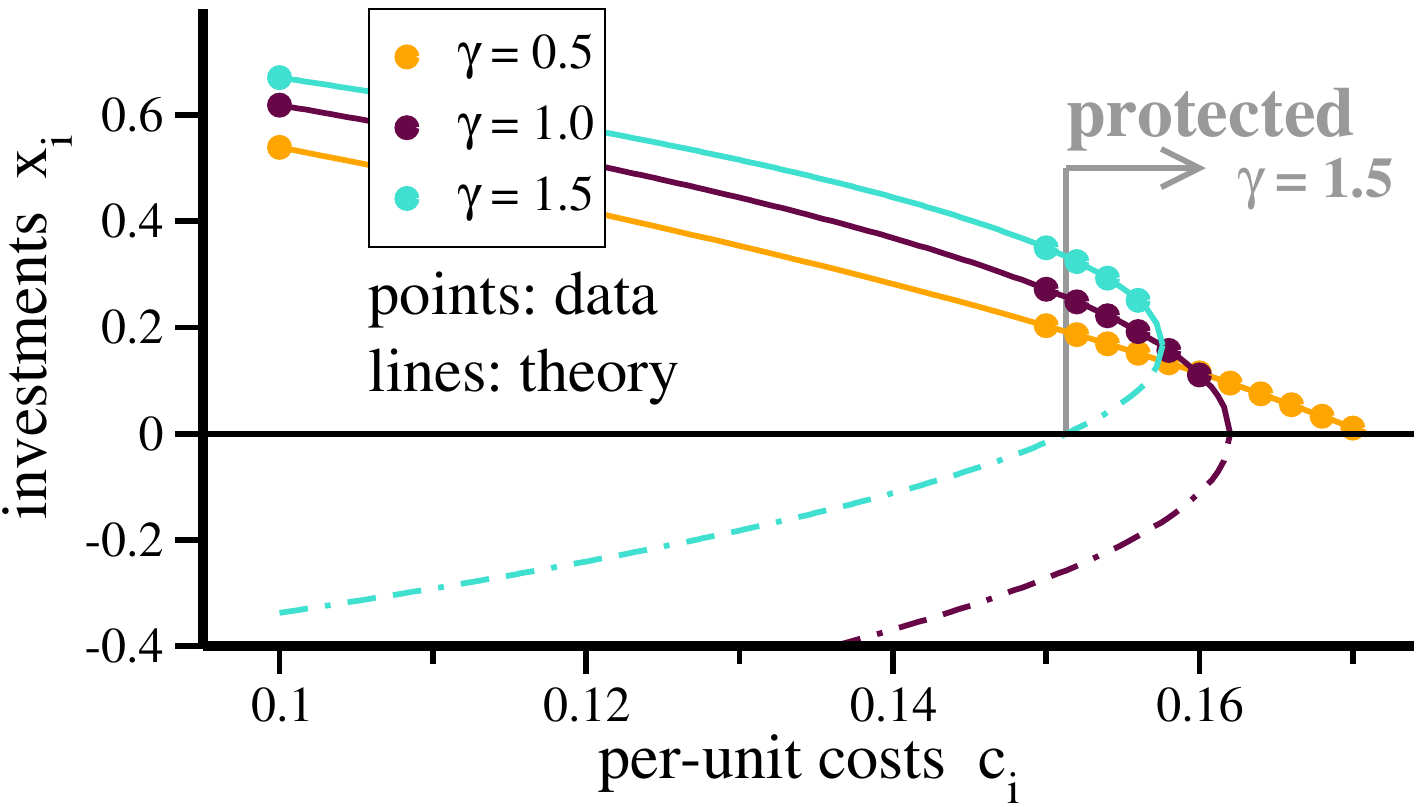}
           }
\caption{{\bf Protected exploitation.}
Results for a logarithmic cost function, as
specified by Eq.~(\ref{CC_gradient}),
with $\gamma\ge0$ encoding how fast marginal
cost falls when investments are increased.
The data has been obtained by adapting the 
individual $x_i$ along the gradient 
$E_i'=dE_i/dx_i$ until convergence 
(points). The solid line corresponds to
the positive roots of the stationarity 
condition (\ref{CC_x_i_roots}). For $\gamma=1.0/1.5$ 
the respective negative roots are also
shown (dash-dotted lines). For $\gamma=1.5$ a 
dynamical entry barrier exists in the region 
indicated by the grey arrow, as explained
further in Fig.~\ref{fig_flow_00}.
As for Fig.~\ref{fig_numerical_00}, cost multipliers
$c_i=0.15+i \Delta c$ have been used, with 
$\Delta c=0.002$. In addition a single oligarch
at $c_i=0.1$ has been included. After decimation $N=12/7/5$
agents remain for $\gamma=0.5/1.0/1.5$.
\label{fig_gamma_00}
}
\end{figure}

Results from numerical simulations 
together with (\ref{CC_x_i_roots}) are 
presented in Fig.~\ref{fig_gamma_00}. 
For $\gamma<1$ one finds that $x_i$ crosses 
zero with a finite slope. This is equivalent 
to the linear behavior observed for the original
model, compare Eq.~(\ref{TOC_gradient_zero}). All
considerations regarding the occurrence of
catastrophic poverty remain hence valid 
for convex and moderately concave cost 
functions, viz.\ for $\gamma<1$. We call this
region the CP phase (CP for catastrophic poverty).

\subsection{Protected exploitation}

By definition, $c_{\rm max}$ is the
locus at which the gradient of the 
individual payoffs (\ref{CC_gradient}) 
changes sign when the individual
investments are vanishing small,
\begin{equation}
\frac{dE_i}{dx_i}\Big|_{x_i=0} = c_{\rm max}-c_i\,.
\label{CC_gradient_0}
\end{equation}
As shown in Fig.~\ref{fig_gamma_00}, agents 
with a cost multiplier $c_i>c_{\rm max}$ may 
survive nevertheless the decimation process 
when their individual investments $x_i$ are
substantial. This is, as discussed above,
the telltale sign of an entry barrier.
Existing agents are hence protected against 
new competitors when $c_i>c_{\rm max}$. 
There is however a limit. Regardless of
the size of their investments, agents will not
survive when cost factors become too large.
This happens when the stationarity condition
$E_i'=0$ cannot be fulfilled any more.

We denote with $c_{\rm node}$ the point
at which the two roots for $x_i$ given 
by (\ref{CC_x_i_roots}) merge. One finds
\begin{equation}
c_{\rm node} = c_{\rm max} +
c_{\rm max} \frac{(\gamma-1)^2}{4\gamma}
= c_{\rm max} \frac{(\gamma+1)^2}{4\gamma}\,.
\label{CC_C_node}
\end{equation}
No solution for $x_i$ exists for $c_i>c_{\rm node}$,
which marks hence the upper bound of the 
protected area. 
As expected, one has that $c_{\rm node}$ 
coincides with $c_{\rm max}$ for $\gamma=1$
and that $c_{\rm node}$ diverges for
$\gamma\to0$. In terms of dynamical systems 
theory, $c_{\rm node}$ corresponds to a
saddle-node bifurcation \cite{gros2015complex}.

\subsection{Forced market exits}

\begin{figure}[!t]
\centerline{
\includegraphics[width=0.75\columnwidth]{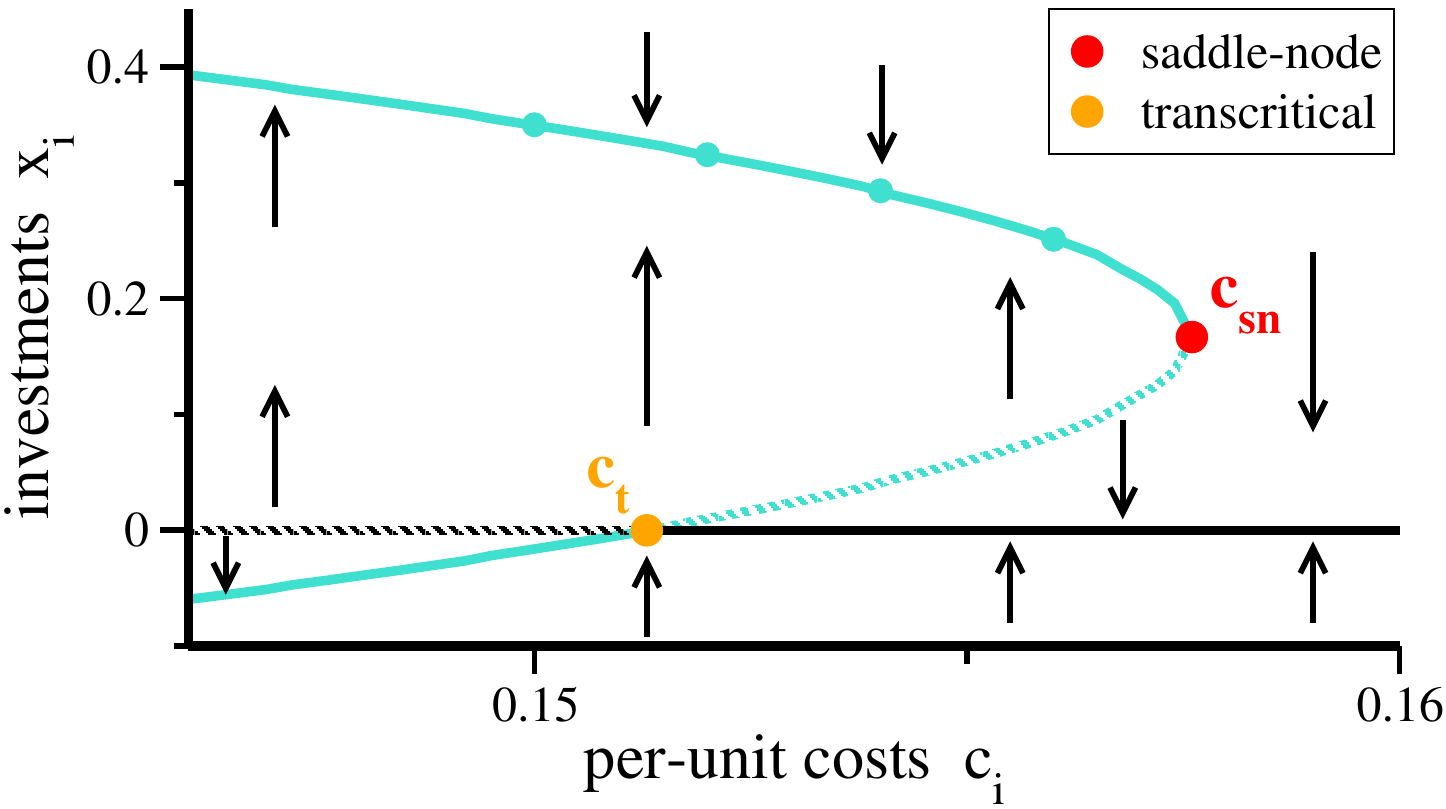}
           }
\caption{{\bf Frozen bifurcation diagram.}
An enlargement of Fig.~\ref{fig_gamma_00}, showing
the flow for $\gamma=1.5$ when $c_{\rm max}=\exp(-x_{\rm tot})$
is kept constant (frozen). The arrows indicate the 
direction of the gradients $E_i'=dE_i/dx_i$
of the payoffs (the flow, in dynamical systems
terminology), as given by (\ref{CC_gradient}).
Solid/shaded lines correspond to stable/unstable
solutions of $E_i'=0$. At the saddle-node bifurcation 
(red filled circle) occurring at $c_{\rm node}$, the
stable and the unstable manifolds merge, corresponding 
respectively to the positive/negative roots defined 
in (\ref{CC_x_i_roots}). The unstable manifold 
undergoes a transcritical transition (orange filled
circle) at $c_{\rm max}$, when crossing the $x_i=0$
line. For $c_i>c_{\rm max}$ the $x_i=0$ line is 
stable, which means that there is a dynamical barrier 
for newcomers to enter the market. 
\label{fig_flow_00}
}
\end{figure}

In order to understand what happens beyond 
the protected area we present in 
Fig.~\ref{fig_flow_00} the frozen bifurcation 
diagram for $\gamma=1.5$. It is given by the 
flow $E_i'$ under the constraint that $x_{\rm tot}=\sum_j x_j$ 
is kept constant, viz.\ frozen. This approximation 
can be used to examine the stability of the 
nullkline $E_i'=0$ in the vicinity of the Nash
equilibrium. Further away, when individual 
investments $x_i$ deviate substantially from 
their stationary value, the frozen approximation 
will however break down.
Together with the saddle-node bifurcation 
at $c_i=c_{\rm node}$ one observes
in Fig.~\ref{fig_flow_00} that the negative root 
of (\ref{CC_x_i_roots}) crosses $x_i=0$ at 
$c_{\rm max}$ via a transcritical bifurcation. 
The later expresses that the nullkline $x_i=0$ 
switches stability at $c_i=c_{\rm max}$, as 
given by (\ref{CC_gradient_0}).

We now examine what happens if a selection of the 
efficient agents further improve their cost 
functions. We assume that this happens slowly, 
such that the system remains all the time close 
to the Nash equilibrium. One is then in a
quasi-stationary state. At the same time the cost
multiplier of the least efficient surviving 
agent, denoted here by $c_N$, is assumed to 
be constant. Both $c_{\rm max}$ and
$c_{\rm node}$ decrease when some agents
lower their $c_i$ progressively, here for $i<N$. 
Eventually, the least efficient agent will be 
squeezed out of the market. The dynamics of 
this forced market exit differs qualitatively for
$\gamma<1$ and $\gamma>1$, viz.\ as a function
of whether entry barriers are present.

For $\gamma<1$ catastrophic poverty is present
and the optimal investment $x_N$ of the least
efficient agent vanishes linearly, 
$\sim(c_{\rm max}-c_N)$, when $c_{\rm max}$
decreases. The point of the market exit
can hence be predicted by analyzing the
rate at which $x_N$ deceases.

For $\gamma>1$ the optimal investment $x_N$
does not vanish when $c_N$ approaches
$c_{\rm node}$. Right at the saddle-node
bifurcation the optimal investment $x_{\rm node}$
is finite, which can be seen by
inserting (\ref{CC_C_node}) into
(\ref{CC_x_i_roots}). Analytically a
square-root singularity exist, namely
that $x_N-x_{\rm node}\sim \sqrt{c_{\rm node}-c_N}$.
Agents will however not be able to resolve
this singular dependence under realistic
market condition involving substantial scattering
and noise. The least efficient investor will
hence not be able to predict its own market
exit by observing $x_N$. The forced market
exit occurs in this sense by a sudden death.

At first sight, the picture developed here 
may seem inconsistent. If $x_{\rm node}$
is finite it follows that agents with
$c_i=c_{\rm node}$ make small but finite 
profits, which can be verified by evaluating
the payoff at the point of the bifurcation.
There is hence no incentive for the least
efficient agent to stop investing into the
common resource. What happens is more subtle.
The flow becomes negative when $c_N$ exceeds 
$c_{\rm node}$ by any amount, as evident
from the frozen phase diagram presented in
Fig.~\ref{fig_flow_00}. The value of $x_N$
decreases for any $c_N>c_{\rm node}$, which
leads in turn to a reduced contribution of
the least efficient investor to $x_{\rm tot}$.
In response, the remainder agents will increase
their own investments, which will further 
reduce $c_{\rm max}$, and consequently also
the optimal $x_N$. In the end, $x_N$ will 
drop to zero.

Effectively, the sudden death forced market exit
results from a collective reorganization of the
market. The agents adapt collectively their
optimal investment, which occurs successively.
The time scale of the `sudden death' forced exit 
is hence determined by the typical time needed to
reach a new Nash equilibrium. The exit is `sudden'
only insofar that the agent cannot predict the
starting point.

\section{Discussion}

Based on a concept proposed  by William Forster 
Lloyd in the 19th century, Garrett Hardin examined 
in 1968 the impact of increased exploitation 
pressures on the state of global natural 
resources \cite{hardin1968tragedy}. An
exploration of a specific mathematical modeling 
framework for the possible outcome, the tragedy 
of the commons (TOC), was however not performed 
at this point. At the center of attention for 
the TOC are common-pool resources for which access 
is unrestricted and free of charges. However, 
free access does not imply that utilization will 
not entail monetary costs. E.g., for Lloyd's
classical example, an open-to-all pasture of 
a village, peasants need to invest first in
their respective livestocks. One needs animals, 
say cows, in order to benefit from a common 
grazing ground. 

For an experimental setup, Ostrom
\cite{ostrom1999coping} used a model 
functionally identical to (\ref{PC_E_i}).
Agents are assumed to dispose of an asset
that can be invested either into a default
activity, like working for an employer and 
getting paid, or into a common pool of resources. 
Assets not used for the default activity
correspond to opportunity costs 
\cite{nicholson2012microeconomic}, which
implies that they are functionally equivalent 
to investment costs. 
It is hence reasonable to assume, as
done throughout this paper, that agents
need to invest when accessing a common
pool of resources. An equivalent observation
holds for certain classes of managed commons. 
An example would be the case of a taxed fishing 
ground. The incurring costs would be equivalent 
to investment costs if taxes are proportional to the 
capacity of the individual boats, but otherwise 
independent of the actual size of the catch.

Lloyd \& Hardin suggested that `the more the 
better' would be the optimal strategy for the 
independently acting agent whenever access to
resources is not restricted, viz.\ when the commons 
are left unmanaged \cite{hardin1968tragedy}. 
If at all, this can however be the case only 
when investment costs are non existent: the 
individual agent will stop increasing the
size of its investment at least once the 
productivity of the commons falls below 
its own per-unit investment cost.
Exploitation rates do not need to diverge however 
even when investment costs are negligibly small.
We showed that large-scale extraction of a common 
resource will be limited when the productivity 
of the commons degrades exponentially with total 
investment $x_{\rm tot}$. This statement holds whenever 
the number $N$ of participating agents remains finite, 
viz.\ when $N<\infty$. However, runaway exploitation occurs 
for vanishing investment costs when the environment degrades 
not exponentially, but as a power law, 
see (\ref{run_condition}). This case could 
hence be considered, in the sense of Lloyd \& Hardin,
to correspond to the paradigmal TOC scenario.

A particular focus of our investigations regards
the dependence of the TOC Nash equilibrium on
group size, viz.\ on $N$. A consistent finding is
that smaller groups will enjoy, ceteris paribus,
larger average payoffs. In the extreme case of
runaway exploitation, as defined by
(\ref{run_condition}), payoffs drop to zero
when the group size exceeds a certain
threshold, remaining finite below.

The effect of group size has been studied 
intensively in game-theoretical studies \cite{nowak2006evolutionary}, 
in particular with regard to the emergence 
of collaboration \cite{nosenzo2015cooperation}.
Of relevance are in this context in addition
psychological and moral components \cite{isaac1988group,yang2013nonlinear,capraro2021mathematical},
like trust and the decoy effect \cite{kumar2020evolution,wang2018exploiting},
longer-term perspectives \cite{santos2011risk},
collective and feedback effects \cite{perc2017statistical,hilbe2018evolution,chen2018punishment,jusup2022social},
and
evolutionary drives \cite{killingback2006evolution,levin2014public}.
A main focus of these and other studies of the
TOC regards the circumstances under which 
over-exploitation will be avoided. Possible
solutions include property management
\cite{libecap2009tragedy}, enhanced
social reputation \cite{milinski2002reputation},
buffer zones \cite{muller2017j}, social 
diversity \cite{santos2008social}, or
community-based institutions in general
\cite{faysse2005coping}. For a general
analysis see \cite{frischmann2013two}.

Here we proposed an alternative view,
pointing out that the steady-state of 
the pure TOC problem is highly non-trivial
already by itself. Commons with unrestricted
access are subject to only modest levels of 
exploitation when group sizes are small, 
and/or when investment costs are substantial. 
The problem becomes severe however when the
investment costs of a large number of 
self-centered investors are small, which
can be expected to be the case for 
technologically advanced societies.
The majority of agents will necessarily 
suffer from substantially reduced incomes 
whenever group sizes are large, scaling
as $(1/N)^2$. This phenomenon is called
`catastrophic poverty'.

Exploitation does not need to be massive for
catastrophic poverty to emerge, which implies
that the status of the commons is not the 
only determinant. 
The steady-state size of the total investment 
$x_{\rm tot}$ may be high, but whether this value
would be sustainable for a specific real-world
pool of resources, or not, this question 
transcends the modeling framework used here.
Instead, agents increase their investments
until the majority of the population 
is close to a self-consistently determined
profitability threshold, with the distance to the 
profitability threshold scaling as $1/N$ in 
equilibrium. A limited number of agents with 
reduced investment costs, the oligarchs, 
may be present in addition. Oligarchs 
escape catastrophic poverty.

Previous studies focused mostly, but not
exclusively \cite{santos2008social}, on agents 
with identical characteristics
\cite{frischmann2019retrospectives}. The
variability between the individual actors
played then a non-essential role. Catastrophic 
poverty will occur also in this limit, viz.\ 
for the case of identical agents. The 
dependency of the Nash equilibrium on the
differences between the individual agents
is nevertheless an integral part of our
investigations. To be specific, agents
have individual per-unit investment 
costs $c_i$. The functional relation between
$c_i$ and the steady-state individual 
payoff, $E(c_i)$, is key for the understanding
of catastrophic poverty. We showed that
this relation, the dispersion relation, is
strictly quadratic, 
$E(c_i) = (c_{\rm max}-c_i)^2/c_{\rm max}$,
where $c_{\rm max}$ is the profitability
threshold. Profits vanish hence quadratically
close $c_{\rm max}$, and not linearly,
which is the cause of catastrophic poverty.

Agents will exploit a common pool of resources 
only when making a net gain. An intriguing finding 
is that the expected number of actively harvesting
agents increases with decreasing investment costs. 
This implies that a given resource will be exploited more 
severely when improved technologies are introduced.
It may explain in reverse why self-organized governance
of the commons is often found in communities with 
somewhat limited access to modern 
technologies \cite{ostrom2009general}.

Our findings hold for a large
class of TOC-models. Of minor relevance for the
occurrence of catastrophic poverty is the
specific dependency of the productivity of
the commons on total investment. Regarding 
the structure of investments costs, catastrophic 
poverty is present for convex and moderately concave 
cost functions. A collective reorganization of 
the market takes however place for strongly concave 
cost functions. In this phase, catastrophic poverty 
is absent. Instead, one finds that entry barriers
for prospective new investors emerge and that the
dynamics of forced market exits changes on a
qualitative level.


\vskip.5pc
\enlargethispage{10pt}
\ack{We thank Roser Valent\'i and Daniel Gros for 
reading the manuscript.}



\begin{thebibliography}{10}

\bibitem{rankin2007tragedy}
Daniel~J Rankin, Katja Bargum, and Hanna Kokko.
\newblock The tragedy of the commons in evolutionary biology.
\newblock {\em Trends in ecology \& evolution}, 22(12):643--651, 2007.

\bibitem{frischmann2019retrospectives}
Brett~M Frischmann, Alain Marciano, and Giovanni~Battista Ramello.
\newblock Retrospectives: Tragedy of the commons after 50 years.
\newblock {\em Journal of Economic Perspectives}, 33(4):211--28, 2019.

\bibitem{ostrom1990governing}
Elinor Ostrom.
\newblock {\em Governing the commons: The evolution of institutions for
  collective action}.
\newblock Cambridge university press, 1990.

\bibitem{feeny1990tragedy}
David Feeny, Fikret Berkes, Bonnie~J McCay, and James~M Acheson.
\newblock The tragedy of the commons: Twenty-two years later.
\newblock {\em Human Ecology}, 18(1), 1990.

\bibitem{cohen1995population}
Joel~E Cohen.
\newblock Population growth and earth's human carrying capacity.
\newblock {\em Science}, 269(5222):341--346, 1995.

\bibitem{battersby2017news}
Stephen Battersby.
\newblock News feature: Can humankind escape the tragedy of the commons?
\newblock {\em Proceedings of the National Academy of Sciences}, 114(1):7--10,
  2017.

\bibitem{faysse2005coping}
Nicolas Faysse.
\newblock Coping with the tragedy of the commons: Game structure and design of
  rules.
\newblock {\em Journal of Economic Surveys}, 19(2), 2005.

\bibitem{hilbe2013evolution}
Christian Hilbe, Martin~A Nowak, and Karl Sigmund.
\newblock Evolution of extortion in iterated prisoner's dilemma games.
\newblock {\em Proceedings of the National Academy of Sciences},
  110(17):6913--6918, 2013.

\bibitem{stewart2014collapse}
Alexander~J Stewart and Joshua~B Plotkin.
\newblock Collapse of cooperation in evolving games.
\newblock {\em Proceedings of the National Academy of Sciences},
  111(49):17558--17563, 2014.

\bibitem{carrozzo2021tragedy}
Alessio Carrozzo~Magli, Pompeo Della~Posta, and Piero Manfredi.
\newblock The tragedy of the commons as a prisoner's dilemma. its relevance for
  sustainability games.
\newblock {\em Sustainability}, 13(15):8125, 2021.

\bibitem{killingback2006evolution}
Timothy Killingback, Jonas Bieri, and Thomas Flatt.
\newblock Evolution in group-structured populations can resolve the tragedy of
  the commons.
\newblock {\em Proceedings of the Royal Society B: Biological Sciences},
  273(1593):1477--1481, 2006.

\bibitem{gore2009snowdrift}
Jeff Gore, Hyun Youk, and Alexander Van~Oudenaarden.
\newblock Snowdrift game dynamics and facultative cheating in yeast.
\newblock {\em Nature}, 459(7244):253--256, 2009.

\bibitem{kummerli2010molecular}
Rolf K{\"u}mmerli and Sam~P Brown.
\newblock Molecular and regulatory properties of a public good shape the
  evolution of cooperation.
\newblock {\em Proceedings of the National Academy of Sciences},
  107(44):18921--18926, 2010.

\bibitem{milinski2008collective}
Manfred Milinski, Ralf~D Sommerfeld, Hans-J{\"u}rgen Krambeck, Floyd~A Reed,
  and Jochem Marotzke.
\newblock The collective-risk social dilemma and the prevention of simulated
  dangerous climate change.
\newblock {\em Proceedings of the National Academy of Sciences},
  105(7):2291--2294, 2008.

\bibitem{capstick2013public}
Stuart~Bryce Capstick.
\newblock Public understanding of climate change as a social dilemma.
\newblock {\em Sustainability}, 5(8):3484--3501, 2013.

\bibitem{bauch2003group}
Chris~T Bauch, Alison~P Galvani, and David~JD Earn.
\newblock Group interest versus self-interest in smallpox vaccination policy.
\newblock {\em Proceedings of the National Academy of Sciences},
  100(18):10564--10567, 2003.

\bibitem{galvani2007long}
Alison~P Galvani, Timothy~C Reluga, and Gretchen~B Chapman.
\newblock Long-standing influenza vaccination policy is in accord with
  individual self-interest but not with the utilitarian optimum.
\newblock {\em Proceedings of the National Academy of Sciences},
  104(13):5692--5697, 2007.

\bibitem{brown2007durability}
Sam~P Brown and Fran{\c{c}}ois Taddei.
\newblock The durability of public goods changes the dynamics and nature of
  social dilemmas.
\newblock {\em PLoS One}, 2(7):e593, 2007.

\bibitem{weitz2016oscillating}
Joshua~S Weitz, Ceyhun Eksin, Keith Paarporn, Sam~P Brown, and William~C
  Ratcliff.
\newblock An oscillating tragedy of the commons in replicator dynamics with
  game-environment feedback.
\newblock {\em Proceedings of the National Academy of Sciences},
  113(47):E7518--E7525, 2016.

\bibitem{tilman2020evolutionary}
Andrew~R Tilman, Joshua~B Plotkin, and Erol Ak{\c{c}}ay.
\newblock Evolutionary games with environmental feedbacks.
\newblock {\em Nature communications}, 11(1):1--11, 2020.

\bibitem{dietz2003struggle}
Thomas Dietz, Elinor Ostrom, and Paul~C Stern.
\newblock The struggle to govern the commons.
\newblock {\em science}, 302(5652):1907--1912, 2003.

\bibitem{ostrom1999coping}
Elinor Ostrom.
\newblock Coping with tragedies of the commons.
\newblock {\em Annual review of political science}, 2(1):493--535, 1999.

\bibitem{nicholson2012microeconomic}
Walter Nicholson and Christopher~M Snyder.
\newblock {\em Microeconomic theory: Basic principles and extensions}.
\newblock Cengage Learning, 2012.

\bibitem{gros2015complex}
Claudius Gros.
\newblock {\em Complex and Adaptive Dynamical Systems, a Primer}.
\newblock Springer, 2015.

\bibitem{landee2014gentle}
Christopher~P Landee and Mark~M Turnbull.
\newblock A gentle introduction to magnetism: Units, fields, theory, and
  experiment.
\newblock {\em Journal of Coordination Chemistry}, 67(3):375--439, 2014.

\bibitem{colombo2021solid}
Luciano Colombo.
\newblock {\em Solid State Physics: A Primer}.
\newblock IoP Publishing, 2021.

\bibitem{lee2017games}
Joung-Hun Lee, Marko Jusup, and Yoh Iwasa.
\newblock Games of corruption in preventing the overuse of common-pool
  resources.
\newblock {\em Journal of Theoretical Biology}, 428:76--86, 2017.

\bibitem{delis2014estimation}
Manthos Delis, Maria Iosifidi, and Efthymios~G Tsionas.
\newblock On the estimation of marginal cost.
\newblock {\em Operations Research}, 62(3):543--556, 2014.

\bibitem{bain1954economies}
Joe~S Bain.
\newblock Economies of scale, concentration, and the condition of entry in
  twenty manufacturing industries.
\newblock {\em The American Economic Review}, 44(1):15--39, 1954.

\bibitem{schmalensee1981economies}
Richard Schmalensee.
\newblock Economies of scale and barriers to entry.
\newblock {\em Journal of political Economy}, 89(6):1228--1238, 1981.

\bibitem{hardin1968tragedy}
Garrett Hardin.
\newblock The tragedy of the commons: the population problem has no technical
  solution; it requires a fundamental extension in morality.
\newblock {\em Science}, 162(3859):1243--1248, 1968.

\bibitem{nowak2006evolutionary}
Martin~A Nowak.
\newblock {\em Evolutionary dynamics: exploring the equations of life}.
\newblock Harvard university press, 2006.

\bibitem{nosenzo2015cooperation}
Daniele Nosenzo, Simone Quercia, and Martin Sefton.
\newblock Cooperation in small groups: the effect of group size.
\newblock {\em Experimental Economics}, 18(1):4--14, 2015.

\bibitem{isaac1988group}
R~Mark Isaac and James~M Walker.
\newblock Group size effects in public goods provision: The voluntary
  contributions mechanism.
\newblock {\em The Quarterly Journal of Economics}, 103(1):179--199, 1988.

\bibitem{yang2013nonlinear}
Wu~Yang, Wei Liu, Andr{\'e}s Vi{\~n}a, Mao-Ning Tuanmu, Guangming He, Thomas
  Dietz, and Jianguo Liu.
\newblock Nonlinear effects of group size on collective action and resource
  outcomes.
\newblock {\em Proceedings of the National Academy of Sciences},
  110(27):10916--10921, 2013.

\bibitem{capraro2021mathematical}
Valerio Capraro and Matja{\v{z}} Perc.
\newblock Mathematical foundations of moral preferences.
\newblock {\em Journal of the Royal Society interface}, 18(175):20200880, 2021.

\bibitem{kumar2020evolution}
Aanjaneya Kumar, Valerio Capraro, and Matja{\v{z}} Perc.
\newblock The evolution of trust and trustworthiness.
\newblock {\em Journal of the Royal Society Interface}, 17(169):20200491, 2020.

\bibitem{wang2018exploiting}
Zhen Wang, Marko Jusup, Lei Shi, Joung-Hun Lee, Yoh Iwasa, and Stefano
  Boccaletti.
\newblock Exploiting a cognitive bias promotes cooperation in social dilemma
  experiments.
\newblock {\em Nature communications}, 9(1):1--7, 2018.

\bibitem{santos2011risk}
Francisco~C Santos and Jorge~M Pacheco.
\newblock Risk of collective failure provides an escape from the tragedy of the
  commons.
\newblock {\em Proceedings of the National Academy of Sciences},
  108(26):10421--10425, 2011.

\bibitem{perc2017statistical}
Matja{\v{z}} Perc, Jillian~J Jordan, David~G Rand, Zhen Wang, Stefano
  Boccaletti, and Attila Szolnoki.
\newblock Statistical physics of human cooperation.
\newblock {\em Physics Reports}, 687:1--51, 2017.

\bibitem{hilbe2018evolution}
Christian Hilbe, {\v{S}}t{\v{e}}p{\'a}n {\v{S}}imsa, Krishnendu Chatterjee, and
  Martin~A Nowak.
\newblock Evolution of cooperation in stochastic games.
\newblock {\em Nature}, 559(7713):246--249, 2018.

\bibitem{chen2018punishment}
Xiaojie Chen and Attila Szolnoki.
\newblock Punishment and inspection for governing the commons in a
  feedback-evolving game.
\newblock {\em PLoS computational biology}, 14(7):e1006347, 2018.

\bibitem{jusup2022social}
Marko Jusup, Petter Holme, Kiyoshi Kanazawa, Misako Takayasu, Ivan Romi{\'c},
  Zhen Wang, Sun{\v{c}}ana Ge{\v{c}}ek, Tomislav Lipi{\'c}, Boris Podobnik, Lin
  Wang, et~al.
\newblock Social physics.
\newblock {\em Physics Reports}, 948:1--148, 2022.

\bibitem{levin2014public}
Simon~A Levin.
\newblock Public goods in relation to competition, cooperation, and spite.
\newblock {\em Proceedings of the National Academy of Sciences}, 111(Supplement
  3):10838--10845, 2014.

\bibitem{libecap2009tragedy}
Gary~D Libecap.
\newblock The tragedy of the commons: property rights and markets as solutions
  to resource and environmental problems.
\newblock {\em Australian Journal of Agricultural and Resource Economics},
  53(1):129--144, 2009.

\bibitem{milinski2002reputation}
Manfred Milinski, Dirk Semmann, and Hans-J{\"u}rgen Krambeck.
\newblock Reputation helps solve the 'tragedy of the commons'.
\newblock {\em Nature}, 415(6870):424--426, 2002.

\bibitem{muller2017j}
Marc~F M{\"u}ller, Mich{\`e}le~C M{\"u}ller-Itten, and Steven~M Gorelick.
\newblock How jordan and saudi arabia are avoiding a tragedy of the commons
  over shared groundwater.
\newblock {\em Water Resources Research}, 53(7):5451--5468, 2017.

\bibitem{santos2008social}
Francisco~C Santos, Marta~D Santos, and Jorge~M Pacheco.
\newblock Social diversity promotes the emergence of cooperation in public
  goods games.
\newblock {\em Nature}, 454(7201):213--216, 2008.

\bibitem{frischmann2013two}
Brett~M Frischmann.
\newblock Two enduring lessons from elinor ostrom.
\newblock {\em Journal of institutional economics}, 9(4):387--406, 2013.

\bibitem{ostrom2009general}
Elinor Ostrom.
\newblock A general framework for analyzing sustainability of social-ecological
  systems.
\newblock {\em Science}, 325(5939):419--422, 2009.

\end{thebibliography}

\end{document}